\documentclass[
journal=jpcbfk,
manuscript=article,
layout=traditional
]{achemso}

\usepackage{siunitx}
\usepackage{graphicx}
\usepackage{dcolumn}
\usepackage{bm}

\usepackage[version=3]{mhchem} 

\author{Peihao Sun}
\email{phsun@stanford.edu}
\altaffiliation{Stanford University Physics Department, 382 Via Pueblo Mall, Stanford, CA 94305, USA}
\author{J. B. Hastings}
\affiliation{SLAC National Accelerator Laboratory, 2575 Sand Hill Rd, Menlo Park, CA 94025, USA}

\author{Daisuke Ishikawa}
\author{Alfred Q. R. Baron}
\affiliation{Materials Dynamics Laboratory, RIKEN SPring-8 Center, 1-1-1 Kouto, Sayo, Hyogo 679-5148, Japan}

\author{Giulio Monaco}
\email{giulio.monaco@unipd.it}
\affiliation{Dipartimento di Fisica e Astronomia, Universit{\`a} di Padova, 35131 Padova, Italy}

\title{Universal Two-Component Dynamics in Supercritical Fluids}

\date{\today}

\begin{document}

\begin{abstract}
Despite the technological importance of supercritical fluids, controversy remains about the details of their microscopic dynamics. In this work, we study four supercritical fluid systems---water, Si, Te, and Lennard-Jones fluid---\emph{via} classical molecular dynamics simulations. A universal two-component behavior is observed in the intermolecular dynamics of these systems, and the changing ratio between the two components leads to a crossover from liquidlike to gaslike dynamics, most rapidly around the Widom line. We find evidence to connect the liquidlike component dominating at lower temperatures with intermolecular bonding, and the component prominent at higher temperatures with free-particle, gaslike dynamics. The ratio between the components can be used to describe important properties of the fluid, such as its self-diffusion coefficient, in the transition region. Our results provide insight into the fundamental mechanism controlling the dynamics of supercritical fluids, and highlight the role of spatiotemporally inhomogenous dynamics even in thermodynamic states where no large-scale fluctuations exist in the fluid.
\end{abstract}

\maketitle 

\section{Introduction}
In the past few decades, supercritical fluids have attracted renewed interest due to their applications in a wide range of chemical and materials processing industries~\cite{Eckert1996}. Most interesting applications of supercritical fluids fall in the region close to the critical point~\cite{Eckert1996, Clifford2000}. There, the fluids exhibit unique properties combining the advantages of liquids (e.g., high densities) and gases (e.g., high diffusivities), and these properties are highly tunable with relatively small changes in temperature, $T$, and pressure, $P$~\cite{Clifford2000}. Thus, it is important to understand these properties and their dependence on the thermodynamic state.

Thanks to many years of research, the thermodynamics of supercritical fluids, which is based on their macroscopic properties, has become well understood. In particular, the concept of the Widom line has been introduced to refer to the line of maxima of a given response function, such as the isobaric heat capacity, $C_P$~\cite{Xu2005}. Although not a rigorous separatrix between liquid and gas states~\cite{Schienbein2018}, the Widom line indicates rapid changes in the thermodynamic properties of supercritical fluids, especially in the near-critical region. Around the Widom line, a crossover between liquidlike and gaslike properties is expected for the fluid~\cite{Gallo2014}.

The picture is less clear when it comes to molecular-scale dynamics of supercritical fluids, which should reveal the microscopic mechanism behind many of the macroscopic properties. One of the first systematic studies on this topic was done by Simeoni et al.~\cite{Simeoni2010}. Using classical molecular dynamics (MD) simulations supported by inelastic x-ray scattering (IXS) data, they observed a crossover in the deep supercritical region along an extension of the Widom line. 

Our previous work~\cite{Sun2020a} focused instead at a region close to the critical point, where the Widom line is very clear. We used both IXS measurements and MD simulations to study the intermolecular dynamics of supercritical water in the region $0.9 < P/P_c < 2.3$, $0.6 < T/T_c < 1.2$, where $P_c$ and $T_c$ are the critical pressure and temperature. Contrary to previous approaches~\cite{Simeoni2010,Bencivenga2007a}, we found that the intermolecular dynamics at a given $P,T$ state cannot be consistently described using models developed for liquids, but instead can be decomposed into two components---a high-frequency component associated with the stretching mode between hydrogen-bonded molecules, and a low-frequency component representing free-particle motions. With changing thermodynamic states, it is the \emph{ratio} between the two components that changes, with a rapid crossover observed near the Widom line. However, remnants of both components can be found on either side of the Widom line.

It is natural to ask whether the observed two-component dynamics is specific to water, whose liquidlike dynamics arises from hydrogen bonds, or can be generalized to other supercritical fluids. In this work, we aim at answering this question by studying the potentials representing four different supercritical fluid systems---water, Si, Te, and Lennard-Jones (LJ) fluid---\emph{via} classical MD simulations. Even though these systems have very different interatomic potentials (see the Methods section below), the two-component behavior is universal in their molecular dynamics. Moreover, we find evidence to associate the liquidlike component with the degree of intermolecular bonding, and the gaslike component with dynamics similar to that in an unbonded, free gas state. As in the case of water, a fast change in the ratio between the two components marks the dynamical crossover, but both components exist on either side of the transition. The fraction of the components can also be used to describe transport properties of the fluid, such as its self-diffusion coefficient.

\section{Methods}
\subsection{Simulation details}
In this study, we investigate four fluid systems with different potential models:

\begin{enumerate}
    \item  Water, with the TIP4P/2005 potential~\cite{Abascal2005}. This potential includes a Lennard-Jones (LJ) interaction between oxygen sites and long-range Coulomb force between all charged sites.
    
    \item Si, with the Stillinger-Weber (SW) potential~\cite{Stillinger1985}. This potential include pairwise  interactions as well as three-body interactions, both short-ranged (cut off at \SI{3.771}{\angstrom}). The three-body interaction term favors local tetrahedral ordering.
    
    \item Te, with an analytical bond-order potential (BOP)~\cite{Ward2012}. This potential considers the effect of bond orders, which are functions of the local environments of the atoms, on the bond energy.
    
    \item LJ fluid, with the shifted-force (sf) potential:
    \begin{equation}
        u^\mathrm{sf}(r) =
            \begin{cases}
                u(r) - u(R_c) - (r-R_c) \left. \frac{du(r)}{dr} \right|_{r=R_c}, & \text{if $r<R_c$} \\
                0, & \text{if $r \geq R_c$}
            \end{cases}
    \end{equation}
    where $r$ is the distance between interacting atoms, $R_c$ is the cutoff distance, and
    \begin{equation}
        u(r) = 4\epsilon \left[ \left( \frac{\sigma}{r} \right)^{12} - \left( \frac{\sigma}{r} \right)^6 \right] \\
    \end{equation}
    is the standard 12-6 potential. $\epsilon$ and $\sigma$ are energy and distance units, respectively. Other units for the LJ fluid can be expressed in terms of $\epsilon$, $\sigma$, and the atomic mass $M$. For example, the units for time is $\tau \equiv \sqrt{M\sigma^2/\epsilon}$. In this work, we set $R_c = 2.5\sigma$. 
\end{enumerate}

The MD simulations are carried out using the LAMMPS simulation package~\cite{Plimpton1995}. The simulation box contains \num{2880} molecules for water and \num{4000} atoms for Si, Te, and LJ fluid. We use $NPT$ ensembles, with a Nos\'e-Hoover thermostat and barostat. The damping constants are \SI{1}{ps} for water, Si, and Te, and $1\tau$ for LJ. After equilibration at each $P,T$ state, the simulation is run for \SI{1}{ns} (with \SI{1}{fs} time steps) for water, \SI{0.4}{ns} (with \SI{1}{fs} time steps) for Si, \SI{1}{ns} (with \SI{2.5}{fs} time steps) for Te, and $1000\tau$ (with $0.001\tau$ time steps) for LJ.

\subsection{Critical parameters}
\begin{figure}
    \centering
    \includegraphics{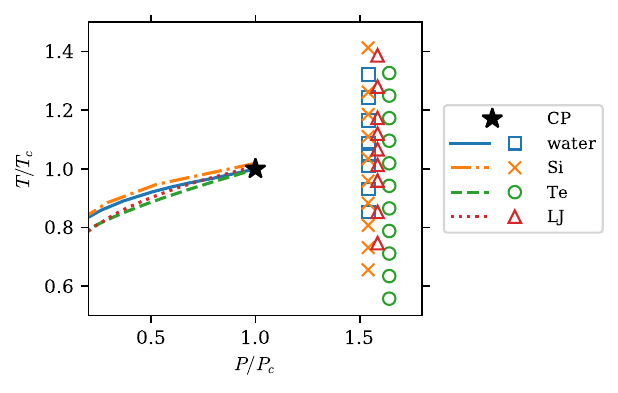}
    \caption{$P$-$T$ phase diagram of the systems in reduced units. The lines show the liquid-vapor coexistence line for the systems (see Refs.~\citenum{Vega2006,Mazhukin2014,Errington2003} for water, Si, and LJ, and Supporting Information for Te), which terminate at the critical point (CP, black star). The symbols show the thermodynamic states included in this study.}
    \label{fig:phase_diagram}
\end{figure}

Table~\ref{tab:CP} presents the critical point parameters for the fluid systems in this study. The TIP4P/2005 model for water, the SW model for Si, and the LJ fluid model are well-studied and their critical parameters can be found in the literature. The critical point parameter for the BOP Te model is determined using a direct MD simulation method~\cite{Alejandre1995}; more details are provided in the Supporting Information. Most of the results below focus on the temperature dependence of the properties of the fluid along an isobar $P \approx 1.6 P_c$; the exact value of $P$ for each system is listed in the last column in Table~\ref{tab:CP}. Figure~\ref{fig:phase_diagram} shows the (reduced) $P$-$T$ phase diagram of all the systems, as well as the thermodynamic states simulated in this study. We note that, as mentioned in the Discussion section below, the two-component phenomenon is \emph{not} an anomaly arising from large-scale critical fluctuations, and the isobars taken are sufficiently away from the critical point. Therefore, the results in this study are robust against errors in the critical point parameters.

\begin{table}
\caption{\label{tab:CP}Critical point parameters for the models used in this study and the isobar pressure $P$. The units for water, Si, and Te are: $T_c$ in \si{K}, $P_c$ and $P$ in \si{bar}, and $\rho_c$ in \si{g/\cubic\cm}. The last column shows the temperature range investigated for each system.}
\centering
\begin{tabular}{cccccc}
\hline\hline
Model & $T_c$ & $P_c$ & $\rho_c$ & $P$ & $T$ range\\
\hline\hline
water~\cite{Vega2011} & $640 \pm 16$ & $146 \pm 7$ & $0.337 \pm 0.008$ & 225 & 546 to 846 \\
\hline
Si~\cite{Makhov2003} & $7925 \pm 250$ & $1850 \pm 400$ & $0.75 \pm 0.10$ & 2850 & 5200 to 11200 \\
\hline
Te & $2080 \pm 40$ & $530 \pm 40$ & $2.17 \pm 0.04$ & 870 & 1160 to 2760 \\
\hline
LJ~\cite{Errington2003} & $0.937 \epsilon/k_B$ & $0.0820 \epsilon/\sigma^3$ & $0.320\sigma^{-3}$ & $0.13 \epsilon/\sigma^3$ & 0.7 to $1.3\,\epsilon/k_B$ \\
\hline\hline
\end{tabular}
\end{table}

\section{Results}
\subsection{Two-component dynamics}
The molecular dynamics of fluids is usually described by the dynamic structure factor, $S(Q, \omega)$, which measures the correlation of density fluctuations in wavenumber ($Q$) and frequency ($\omega$) space~\cite{Boon1991}. It is defined as
\begin{equation} \label{eq:S_def}
    S(Q, \omega) = \frac{1}{2\pi} \int_{-\infty}^{\infty} dt \; e^{-i \omega t} \left< \rho_Q^*(0) \rho_Q(t) \right>,
\end{equation}
where angular brackets indicate the ensemble average, and $\rho_Q(t)=\sum_{n=1}^{N} e^{i{\bf Q}\cdot {\bf r}_n(t)} \left/ \sqrt{N} \right.$ is the density in $Q$-space at time $t$, ${\bf r}_n(t)$ being the position of the $n^\text{th}$ atom. In this paper, we take the \emph{classical} limit. $S(Q, \omega)$ is one of the most important functions to describe the molecular dynamics of fluids, as it contains all the relevant information on the dynamics of the system~\cite{Boon1991}. Moreover, at wavenumbers approaching intermolecular scales ($Q \sim \si{\per\angstrom}$), $S(Q, \omega)$ can be directly measured using inelastic neutron and x-ray scattering~\cite{Boon1991, Baron2020}.

The dynamics in different thermodynamic states can be conveniently compared using the longitudinal current correlation $J(Q, \omega)$, defined by replacing the density $\rho_Q(t)$ in Eq.~(\ref{eq:S_def}) with the longitudinal current $j_{Q, l} (t) = \sum_{n=1}^{N} v_{n,l}(t) e^{i{\bf Q}\cdot {\bf r}_n(t)}\left/ \sqrt{N} \right.$. Here, $v_{n,l}(t)$ denotes the velocity of the $n^\text{th}$ atom along the direction of ${\bf Q}$. It bears a simple relation to $S(Q, \omega)$~\cite{Boon1991}:
\begin{equation} \label{eq:JL_S_relation}
    J_l(Q, \omega) = \frac{\omega^2}{Q^2} S(Q, \omega).
\end{equation}
$J(Q, \omega)$ obeys the classical sum rule~\cite{Boon1991}:
\begin{equation} \label{eq:sum_rule}
    \frac{M}{k_B T} \int_{-\infty}^\infty J_l(Q, \omega) d\omega = 1,
\end{equation}
where the pre-factor contains only the molecular mass $M$, the Boltzmann constant $k_B$, and the temperature $T$, all of which are known constants for the simulation. This provides a simple way to normalize and compare the spectra for different thermodynamic states.

\begin{figure}
    \centering
    \includegraphics[width=3.5in]{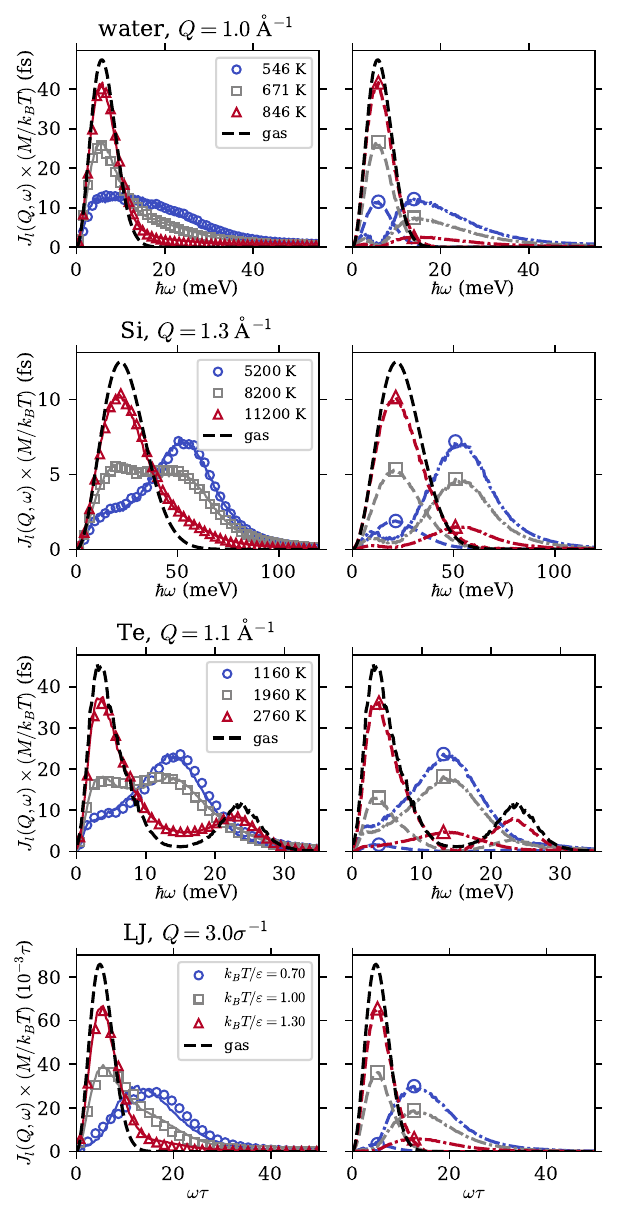}
    \caption{Two-component behavior in the longitudinal current correlation function. Each row shows a different fluid system. The left columns show data from simulation (symbols) along with the NMF fit (solid lines). For each system we choose three states along an isobar $P \approx 1.6 P_c$: a low-temperature, liquidlike state (blue circles), an intermediate state in the crossover region (grey squares), and a high-temperature, gaslike state (red triangles). The right column shows the L component (dash-dotted lines) and G component (dashed lines) obtained from NMF, and the peak positions are marked by corresponding symbols. For reference, we show in both columns the gas limit as black dashed lines without symbols. $\sigma$ and $\epsilon$ are LJ units.}
    \label{fig:JL}
\end{figure}

With the help of this normalization, the two-component behavior in the fluid systems becomes clear. This can be seen in Fig.~\ref{fig:JL}, where the symbols on the left column show the normalized spectra, $J_l(Q, \omega) \times (M/k_B T)$, obtained from MD simulations. Each row presents one of the four fluid systems in this study---water, Si, Te, and LJ fluid---as indicated. For each system, three temperature points are taken along an isobar of $P \approx 1.6 P_c$ as indicated in Table~\ref{tab:CP}: a low-temperature state (blue circles), an intermediate temperature state (grey squares), and a high-temperature state (red triangles). The $Q$ value is chosen to be approximately $0.5 Q_m$, where $Q_m$ is the position of first peak in the structure factor $S(Q)$; in real space, this $Q$ corresponds to approximately twice the average intermolecular distance. We note that the same two-component phenomenon can be observed at other $Q$ values at least in the range from $0.3Q_m$ to $0.8Q_m$, as was also the case in our previous work~\cite{Sun2020a}.

The black lines in Fig.~\ref{fig:JL} show the spectra expected of the gas state. For water, Si, and LJ fluid, this is taken to be the free-particle limit, assuming simply a Maxwell-Boltzmann velocity distribution with no interaction~\cite{Boon1991,Sun2020a}:
\begin{equation} \label{eq:free}
    J_l^\text{free}(Q, \omega) = \frac{\omega^2}{\sqrt{2\pi} Q^3 v_0} \exp\left[ -\frac{1}{2}\left( \frac{\omega}{Q v_0}\right)^2 \right],
\end{equation}
where $v_0 \equiv \sqrt{k_B T/M}$ is the thermal velocity. The temperature is taken to be the same as the high-$T$ state, although small changes in $T$ lead only to a slight shift ($\propto \sqrt{T}$) in the peak position and do not appear to significantly influence the results below. For Te, the gas phase is diatomic (i.e., it consists of Te$_2$ dimers), so there is an additional peak around \SI{23}{meV} corresponding to the dimer stretching mode (see Supporting Information for more details). Hence, a simple expression cannot be obtained for $J_l^\text{free}(Q, \omega)$, and we use instead a low-$P$ spectrum at \SI{100}{bar}, \SI{2760}{K}, where the density is only \SI{0.109}{g/\cubic\cm} compared to the critical density of \SI{2.17}{g/\cubic\cm}. It can be seen that the high-$T$ spectrum is close to the gas state for all systems.

From these plots it is clear, particularly for Si and Te, that the intermediate state contains features of both the low and the high temperature spectra as in the case of water~\cite{Sun2020a}. Specifically, the intermediate spectrum in Si shows both the peak around \SI{55}{meV} which is prominent in the low-$T$ state and the peak around \SI{20}{meV} which dominates the high-$T$ state, and similarly for Te (including the dimer oscillation peak around \SI{23}{meV}). In the case of the LJ fluid, even though we do not observe two distinct peaks, the intermediate temperature spectrum can still be interpreted as a linear combination of the high and low temperature states. In addition, as will be shown below, this interpretation can be used to predict other properties of the LJ fluid in the same way as for the other systems. Therefore, our results show that there is a universal two-component behavior in the supercritical fluids under study.

\subsection{NMF analysis and the liquidlike to gaslike transition}
In order to describe the spectra quantitatively, a method is needed to extract the two components. To our knowledge, however, no existing theory can adequately describe the two-component phenomenon and provide a model to fit the data. Therefore, we adopt the nonnegative matrix factorization (NMF) method~\cite{Hoyer2004} used in our previous study~\cite{Sun2020a}, which provides a model-free way to extract the components in the spectra. Mathematically, we optimize the decomposition 
\begin{equation} \label{eq:NMF_def}
    \frac{M}{k_B T} J_l(Q, \omega; P, T) = c_\text{L}(P, T) J_l^\text{L}(Q, \omega) + c_\text{G}(P, T) J_l^\text{G}(Q, \omega) 
\end{equation}
where $J_l^\text{L}(Q, \omega)$ and $J_l^\text{G}(Q, \omega)$ are the L and G components dominating in the liquidlike (low $T$) and gaslike (high $T$) states, respectively; the shapes of these components are assumed to be independent of $P$ and $T$. The pressure and temperature-dependence of the normalized spectra are captured entirely in the coefficients $c_\text{L}(P, T)$ and $c_\text{G}(P, T)$. When fitting, we include all temperatures along the isobar and add the gas state as well. It also turns out that spectra from different $Q$ can be fit together, resulting in the same coefficients $c_\text{L}(P, T)$ and $c_\text{G}(P, T)$. Because we are interested in molecular-scale dynamics, in this work we typically use data from $0.3 Q_m$ to $0.8 Q_m$, which corresponds to length scales on the same order as the average intermolecular distance. Small changes in the $Q$ range used for fitting do not have a significant influence on the results below.  When $Q<0.3 Q_m$, the data tend to be noisier because of the finite system size and energy resolution.

Results of the NMF decomposition are shown in Fig.~\ref{fig:JL}. On the left column, the solid lines show the NMF fit (sum of the components), which agrees well with data (symbols); on the right column, the G and L components are shown as dashed and dash-dotted lines, respectively, with the corresponding symbols indicating their respective peak positions. In all systems, the G component is close in shape to the gas state spectrum, and the L component peaks at a higher frequency. With increasing temperature, the spectral weight shifts from the L to the G component, leading to a liquidlike to gaslike transition.

Because of the sum rule, Eq.~(\ref{eq:sum_rule}), we normalize the L and G components as well so that $\int_{-\infty}^\infty J_l^\text{L,G}(Q, \omega) d\omega = 1$. As a result, $c_\text{L}+c_\text{G} = 1$, so we may interpret $c_\text{L}$ and $c_\text{G}$ as the \emph{fraction} of the L and G components. 
If we now define the parameter $f\equiv c_\text{L}$, it can be seen from Eq.~(\ref{eq:NMF_def}) that the spectral evolution is captured entirely by the single parameter $f$ as a function $P$ and $T$, and any dynamical crossover on an isobar should show up when plotting $f(T)$.

\begin{figure}
    \centering
    \includegraphics[width=0.7\columnwidth]{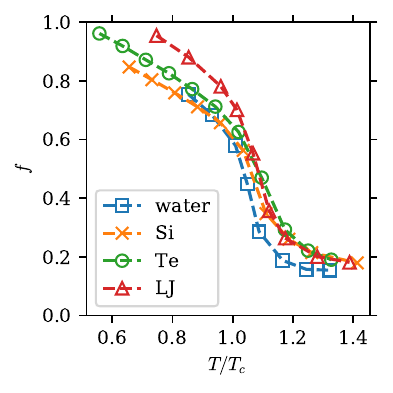}
    \caption{Dependence of the fraction of the L component, $f$, on reduced temperature $T/T_c$ along the isobar $P \approx 1.6 P_c$. Blue squares: water; yellow crosses: Si; green circles: Te; red triangles: LJ fluid. Dashed lines are a guide to the eye.}
    \label{fig:f_vs_T}
\end{figure}

\begin{figure}
    \centering
    \includegraphics[width=0.95\columnwidth]{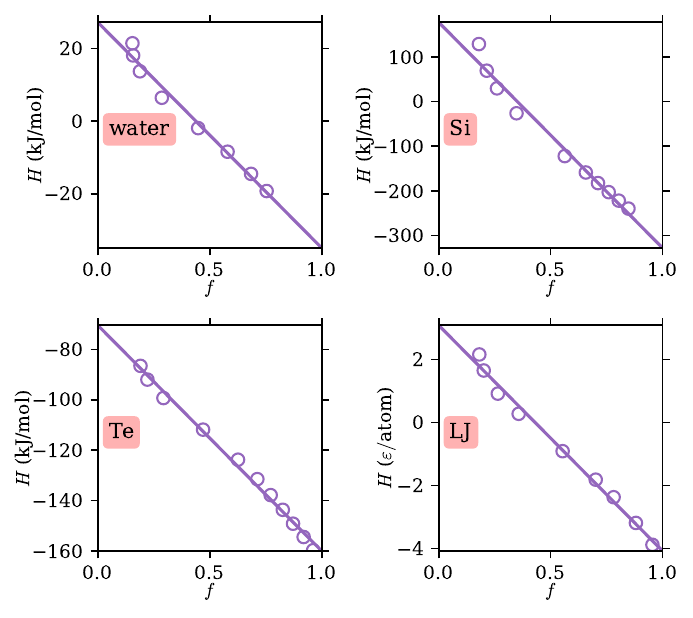}
    \caption{Relation between the enthalpy, $H$, and the parameter $f$. Data are shown as empty circles, and the linear fits as solid lines. $\epsilon$ is the LJ energy unit (see Method).}
    \label{fig:f_vs_H}
\end{figure}

Therefore, in Fig.~\ref{fig:f_vs_T} we present $f$ as a function of reduced temperature $T/T_c$ for all four systems. The overall shape and value of the curves are very similar for all the systems. This is consistent with van der Waals's law of corresponding states~\cite{Pitzer1939} and provides evidence for the universality of the two-component behavior among supercritical fluids. In particular, all curves show an ``S'' shape with a rapid decrease slightly above $T/T_c = 1$.
The position of the fast change in $f$ agrees well with the expected location of the Widom line. To show this, we plot in Fig.~\ref{fig:f_vs_H} the enthalpy, $H$, against the parameter $f$. The former can be easily obtained from MD simulations. An approximately linear relation can be seen between $f$ and $H$ for all systems, with linear fits shown as solid lines. Because the isobaric heat capacity, $C_P$, is the derivative of $H$ with respect to temperature along the isobar, the linearity between $f$ and $H$ implies that $|df/dT|$ peaks at roughly the same temperature as $C_P$, i.e., near the Widom line. In other words, the dynamics change most rapidly around the Widom line. We note that although the Widom line here has a specific definition ($C_P$ maximum along an isobar), in the near-critical region it is expected to lie close to the Widom lines obtained by other definitions as well. For example, in the case of water, it has been shown that the rapid changes in $f$ are close to the Widom lines with several different definitions~\cite{Sun2020a}.

\subsection{The L component and intermolecular bonding}
Having established above that the G component corresponds to the gas state, we now turn to the physical origin of the L component. In the case of water, our previous work~\cite{Sun2020a} has provided evidence that this component is related to the O---O stretching motion between hydrogen-bonded molecules. Therefore, it is reasonable to hypothesize that the L component in the other systems is related to intermolecular bonding as well.

To investigate this, it is necessary to define ``bonding'' for these systems. Because the LJ fluid has only a pairwise interaction that depends solely on the interatomic distance, it is natural to define a cutoff distance $R_b$ below which a pair is considered bonded. In the following, we take $R_b=1.6\sigma$, close to the first minimum in the radial distribution function $g(r)$ in the low-temperature state at $T=0.7\epsilon/k_B$, $P=0.13\epsilon/\sigma^3$ (see Supporting Information for details on $g(r)$).

The cases of Si and Te are in principle more complicated. Unlike water, whose hydrogen bonds can be defined by the geometry and/or the interaction energy between two molecules, Si and Te contain interactions terms that involve three or more atoms (see the Methods section). To our knowledge, there is no established way to define ``bonding'' in these systems. Hence, we simply define two atoms to be bonded if they are closer than a cutoff distance $R_b$. As in the case of the LJ fluid, $R_b$ is chosen to be around the first minimum in the radial distribution function for the low-temperature state, which is about \SI{3.5}{\angstrom} and \SI{4.2}{\angstrom} for Si and Te, respectively (see Supporting Information).

\begin{figure}
    \centering
    \includegraphics[width=0.95\columnwidth]{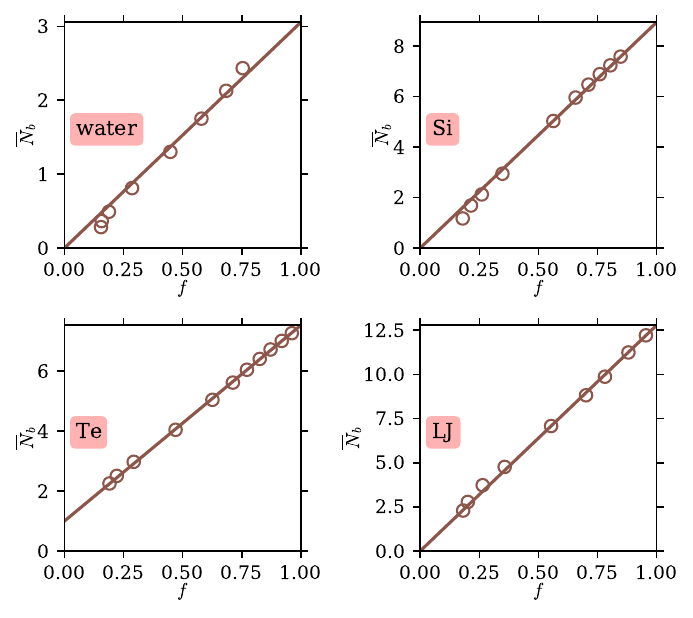}
    \caption{Relation between the number of bonds per atom (or water molecule), $\overline{N}_b$, and the parameter $f$. Data are shown as empty circles. For water, Si, and LJ, the data are consistent with an intercept of $\overline{N}_b=0$ at $f=0$, as the solid lines show. For Te, the data are consistent with an intercept of $\overline{N}_b=1$ at $f=0$ shown by the solid line.}
    \label{fig:f_vs_b}
\end{figure}

In Fig.~\ref{fig:f_vs_b}, the circles show the average number of bonds each atom (or water molecule) has, $\overline{N}_b$, plotted against the parameter $f$. For water, as in our previous work, we use a common definition for hydrogen bonding: two molecules are hydrogen-bonded if their O--O distance is less than \SI{3.5}{\angstrom} and the O$\cdots$O---H angle is less than \ang{30}~\cite{Luzar1996a, Luzar1996}. For Si, Te, and LJ, we use the cutoff distance definition mentioned above. The data show very good linearity between $\overline{N}_b$ and $f$. Moreover, for water, Si, and LJ, the data are consistent with a zero intercept at $f=0$, as the solid lines show. For Te, as mentioned above and shown in more details in the Supporting Information, the gas state consists of Te$_2$ dimers, so we expect each atom to have exactly one bond. Indeed, the data are consistent with an intercept of $\overline{N}_b=1$ at $f=0$, as the solid line shows. These results strongly support that the L component, which dominates in low-temperature, liquidlike states, is directly related to intermolecular bonding for all systems studied. In the gas state, little to no bonding remains, and the L component disappears. We note that for water, using other hydrogen bonding definitions with various levels of strictness does not alter the conclusion, and for Si, Te, and LJ, the conclusion is robust against changes in the cutoff distance being used up to at least 10\% (see Supporting Information).

\subsection{Application: modeling the self-diffusion coefficient}
Our results above have provided evidence for the two-component dynamical behavior and have shown that $f$ is a descriptor for the microscopic dynamics in the liquidlike to gaslike crossover. Since the microscopic dynamics is closely related to macroscopic transport properties
, there should be a close relation between $f$ and transport properties as well. Below we show one such example.

One of the most important transport properties for supercritical fluids, especially for industrial applications, is the self-diffusion coefficient, $D$. This quantity can be easily obtained from MD simulations using the mean squared displacement~\cite{Boon1991}:
\begin{equation}
    D = \frac{1}{6} \lim_{t \to \infty} \frac{\left< |{\bf r}(t) - {\bf r}(0)|^2\right>}{t},
\end{equation}
where ${\bf r}(t)$ the position of a given particle at time $t$. Here angular brackets denote the ensemble average. For water, we use the position of the O atom. The simulation times are long enough to reach the $t \to \infty$ limit. Alternatively, $D$ can be obtained using the velocity autocorrelation function~\cite{Boon1991}:
\begin{equation}
    D = \frac{1}{3} \int_0^\infty \left< {\bf v}(0) \cdot  {\bf v}(t) \right> dt,
\end{equation}
where ${\bf v} (t)$ is the velocity of a given particle at time $t$. The results from the two methods agree within 5\%.

Earlier work~\cite{Gallo2014} found that the self-diffusion coefficient for supercritical water appeared to follow an Arrhenius equation in the liquidlike and the gaslike region along each isobar. A dynamical crossover was found in between, but no specific model was given to describe it. Here we propose a model in which the parameter $f$ is used to describe this transition. 

A good model should reduce to the observed dependence in the gaslike and liquidlike limits. In the limit of a dilute gas, it is well known~\cite{Slattery1958,Chapman1939} that $D$ has a power-law dependence on either the temperature $T$ or the density $\rho$ ($T$ and $\rho$ are inversely related on an isobar by the ideal gas law). Since in this limit the density should be proportional to the number of bonds per atom, which is in turn proportional to $f$, we expect $D$ to have a power-law dependence on $f$ as well. In the dense liquid limit, $D$ is often described instead by the free volume model~\cite{Doolittle1951,Cohen1959}: $D \propto \exp(-A_v/V_f)$, where $A_v$ is a constant and $V_f$ the free molecular volume (i.e., the average volume per molecule in excess of its Van der Waals volume). Under the framework of our two-component dynamics description, we draw an analogy between the free volume, $V_f$, and the fraction of the gaslike component, $1-f$. This can be justified by noting that, as shown above, the gaslike dynamic component corresponds to free-particle-like diffusive motions in the fluid. Therefore, in the dense liquid limit, we expect $D \propto e^{-A/(1-f)}$ where $A$ is a constant.

\begin{figure}
    \centering
    \includegraphics[width=0.95\columnwidth]{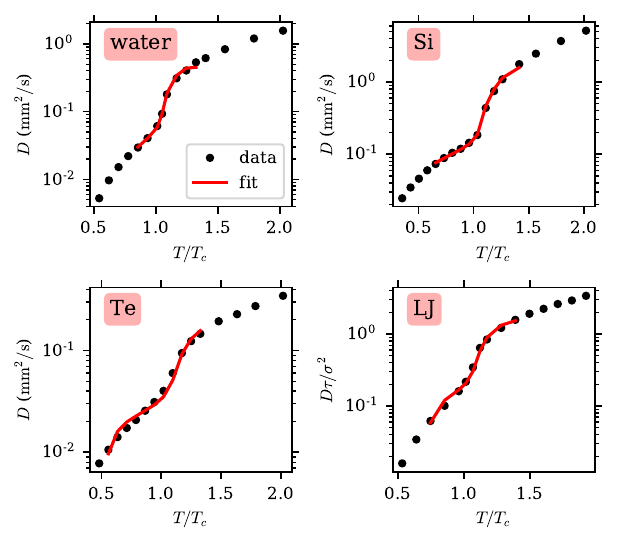}
    \caption{The self-diffusion coefficient, $D$, plotted against reduced temperature along an isobar $P \approx 1.6 P_c$. The MD data are shown as black dots. The red lines show the fits using Eq.~(\ref{eq:D_model}) over the range where $f$ is available through the two-component analysis.}
    \label{fig:D_vs_f}
\end{figure}

Combining the two limits, we build the following model for the self-diffusion coefficient:
\begin{equation} \label{eq:D_model}
    D = D_0 f^{-n} \exp \left( -\frac{A}{1-f} \right),
\end{equation}
where $D_0$, $n$, $A$ are constants. In the gaslike limit, $f \to 0$, so $D \to D_0 f^{-n} e^{-A} \propto f^{-n}$, i.e. it shows the expected power-law dependence. In the liquidlike limit, $f \to 1$, so $D \to D_0 e^{-A/(1-f)}$ in line with the discussion above. We use Eq.~(\ref{eq:D_model}) to fit the data with $D_0$, $n$, and $A$ as fit parameters, and the results are shown in Fig.~\ref{fig:D_vs_f}. The model is able to fit the data well, including the crossover region near the Widom line where $D$ increases rapidly with temperature. Values of the fit parameters are shown in Table~\ref{tab:D_fit_params}. Except for Te, the exponent $n$ in the gas limit is similar to literature values: $n=1.5$ from the Chapman-Enskog theory~\cite{Chapman1939}, and $n=1.823$ for nonpolar systems according to Slattery and Bird's fit for experimental data~\cite{Slattery1958}. Note that, as mentioned above, on an isobar we expect $f \propto \rho \propto T^{-1}$, so we can rewrite expressions in the literature in terms of $f$. Through this example, we show that $f$ can be used to describe macroscopic transport properties across the liquidlike to gaslike transition, connecting the limits of a dense liquid and a dilute gas. Given the proportionality between $f$ and the number of bonds, $\overline{N}_b$, Eq.~(\ref{eq:D_model}) may also be rewritten in terms of $\overline{N}_b$ and expanded to cover a wider range of thermodynamic states. This can be grounds for future investigations.

\begin{table}
\caption{\label{tab:D_fit_params}Fit parameters for the self-diffision coefficient model, Eq.~(\ref{eq:D_model}). $D_0$ is in units of \si{\mm\squared/s} (for water, Si, Te) or $\sigma^2/\tau$ (for LJ).}
\centering
\begin{tabular}{cccc}
\hline
System & $\ln D_0$ & $n$ & $A$ \\
\hline\hline
water & $-3.16 \pm 0.28$ & $1.38 \pm 0.12$ & $0.181 \pm 0.075$ \\
\hline
Si & $-2.66 \pm 0.11$ & $1.84 \pm 0.07$ & $0.034 \pm 0.021$ \\
\hline
Te & $-3.83 \pm 0.08$ & $1.22 \pm 0.08$ & $0.033 \pm 0.006$ \\
\hline
LJ & $-1.93 \pm 0.10$ & $1.42 \pm 0.08$ & $0.044 \pm 0.008$ \\
\hline
\end{tabular}
\end{table}

\section{Discussion \& Conclusions}
In order to demonstrate the universality of the two-component phenomenon, we have chosen in our study four systems containing very different interatomic interactions (see the Methods section for more details)---the simple pairwise LJ potential, TIP4P/2005 water~\cite{Abascal2005} with long-range Coulomb forces, Stillinger-Weber (SW) silicon~\cite{Stillinger1985} with a three-body term favoring local tetrahedral coordination, and tellurium bond-order potential~\cite{Ward2012} where the gas phase is diatomic. The appearance of the two-component dynamics in all systems
shows that this phenomenon is not specific to the local bonding mechanism, but common among several supercritical fluid systems. Consequently, any theory describing the molecular-scale dynamics of supercritical fluids, particularly the crossover between liquidlike and gaslike behavior, should take into account the existence of at least two components in the dynamics.

We note that the two-component phenomenon is not an anomaly arising from large-scale critical fluctuations, since the thermodynamic states in this study are sufficiently far away from the critical point and no such large-scale fluctuations are observed in our simulations. Instead, our results suggest the presence of spatiotemporally heterogeneous dynamics on the \emph{molecular} scale, reflecting unbounded and bounded particle motions. Notably, a recent work~\cite{Ha2018} using machine learning on local structural information has also found the existence of molecular-scale heterogeneities in supercritical LJ fluids. Because of this, the two-component phenomenon is not expected to appear in the long wavelength (low $Q$) limit. This has not been explored in our study by the low $Q$ cutoff around $0.3Q_m$ as mentioned in the Results section. Nonetheless, macroscopic quantities are influenced by their microscopic mechanisms and, as shown above, the use of the two-component model for the molecular dynamics can help build a more fundamental understanding of macroscopic properties such as the diffusion coefficient.

We mention here another dynamical crossover proposed in the literature, the ``Frenkel line'', which separates the supercritical region into ``rigid'' and ``non-rigid'' fluids depending on the relaxation time of the system~\cite{Brazhkin2012}. The underlying assumption there is that a single relaxation time describes the dynamics of all the fluid.  Here we have shown, at least in the near-critical region we have investigated, that the dynamics is spatiotemporally heterogeneous. Thus it is not appropriate to describe the dynamics as purely liquid-like or gas-like, but rather a combination of both. In our previous work on supercritical water~\cite{Sun2020a} including both experimental and simulation results, no significant change was observed near the proposed Frenkel line position. However, we have not investigated the deep supercritical region where the Frenkel line might also exist~\cite{Brazhkin2012}; this may be a subject for future studies.

A limitation of our methodology using the NMF decomposition is the assumption that the shapes of the components do not change with the thermodynamic state. This of course does not work at \emph{all} temperatures and pressures; for example, going to extremely high temperatures, the free-particle limit will be broadened according to Eq.~(\ref{eq:free}). However, the fact that the NMF fit shown in Fig.~\ref{fig:JL} works well indicates that this assumption is valid over the temperature range under study, around $0.6T_c$ to $1.4T_c$. As mentioned in the introduction, this range around the critical point is the most interesting for applications. A more rigorous theory taking into account the change in the shape of the components may be able to describe a wider range of thermodynamic conditions.

We emphasize that one interesting point of our approach is that it can be checked against scattering experiments, for example high-resolution inelastic X-ray scattering~\cite{Baron2020}. These experiments directly measure the dynamical structure factor, $S(Q, \omega)$~\cite{Boon1991,Baron2020}, and the simple relation given by Eq.~(\ref{eq:JL_S_relation}) connects it to $J_l(Q, \omega)$. The $J_l(Q, \omega)$ spectra are all that is needed for the two-component analysis and the extraction of the parameter $f$. Thus, $f$ is a descriptor of microscopic dynamics that is experimentally accessible and, as shown above, it is connected with various other properties of the fluid. In our previous work~\cite{Sun2020a}, we have indeed used inelastic X-ray scattering to measure the molecular dynamics of supercritical water, and found excellent agreement between experimental data and MD simulation results. Similar measurements can be done on other supercritical fluid systems as well to verify experimentally the universality of the two-component phenomenon found in this study. We note that, while the TIP4P/2005 water potential and the LJ potential have been shown to reproduce well experimental data on the dynamics of supercritical water~\cite{Sun2020a} and argon~\cite{Bolmatov2015a}, the Si and Te potentials used in this study have not been optimized or checked against experimental data in the supercritical region, since no data is yet available.

This universality and the close relation between intermolecular bonding and the L component is reminiscent of the well-known lattice gas model~\cite{Yang1952,Lee1952}, which forms the basis connecting the liquid-gas critical point to the 3D Ising universality class. In the lattice gas model, a liquid-to-gas transition takes place with the breaking of bonds, which is similar to the behavior of $f$ and its connection to intermolecular bonding found in our study. Furthermore, we note that both in the lattice gas model and in our two-component analysis the transition from liquidlike to gaslike happens gradually with a continuous loss of bonds. Therefore, our study suggests that the understanding of supercritical fluids based on the lattice gas model may be extended into the description of their molecular dynamics as well.

In conclusion, our results show that the two-component phenomenon in the molecular dynamics, previously observed in supercritical water~\cite{Sun2020a}, is universal among several supercritical fluid systems with different intermolecular interactions. While the gaslike (G) component corresponds to free-particle motion in a dilute gas, the liquidlike (L) component can be associated with intermolecular bonding (a generalization of hydrogen-bonding in the case of water). These observations are shown to have important implications for transport properties such as the self-diffusion coefficient, particularly in bridging the liquidlike to gaslike transition, which is relevant to industrial applications.

\paragraph{Supporting Information.} Details on: i) the dimer gas phase of the Te BOP potential, ii) simulation results to determine the critical parameters of the Te BOP potential, iii) the components retrieved by NMF fit and the dispersion relation, iv) hydrogen bond definitions, and v) the bonding definition for Si, Te, and LJ.

\begin{acknowledgement}
This work is supported by the U.S. Department of Energy, Office of Science, Office of Basic Energy Sciences under Contract No.~DE-AC02-76SF00515. Some of the computing for this project was performed on the Sherlock cluster. We would like to thank Stanford University and the Stanford Research Computing Center for providing computational resources and support that contributed to these research results.
\end{acknowledgement}

\bibliography{main}

\providecommand{\latin}[1]{#1}
\makeatletter
\providecommand{\doi}
  {\begingroup\let\do\@makeother\dospecials
  \catcode`\{=1 \catcode`\}=2 \doi@aux}
\providecommand{\doi@aux}[1]{\endgroup\texttt{#1}}
\makeatother
\providecommand*\mcitethebibliography{\thebibliography}
\csname @ifundefined\endcsname{endmcitethebibliography}
  {\let\endmcitethebibliography\endthebibliography}{}
\begin{mcitethebibliography}{34}
\providecommand*\natexlab[1]{#1}
\providecommand*\mciteSetBstSublistMode[1]{}
\providecommand*\mciteSetBstMaxWidthForm[2]{}
\providecommand*\mciteBstWouldAddEndPuncttrue
  {\def\EndOfBibitem{\unskip.}}
\providecommand*\mciteBstWouldAddEndPunctfalse
  {\let\EndOfBibitem\relax}
\providecommand*\mciteSetBstMidEndSepPunct[3]{}
\providecommand*\mciteSetBstSublistLabelBeginEnd[3]{}
\providecommand*\EndOfBibitem{}
\mciteSetBstSublistMode{f}
\mciteSetBstMaxWidthForm{subitem}{(\alph{mcitesubitemcount})}
\mciteSetBstSublistLabelBeginEnd
  {\mcitemaxwidthsubitemform\space}
  {\relax}
  {\relax}

\bibitem[Eckert \latin{et~al.}(1996)Eckert, Knutson, and
  Debenedetti]{Eckert1996}
Eckert,~C.~A.; Knutson,~B.~L.; Debenedetti,~P.~G. {Supercritical fluids as
  solvents for chemical and materials processing}. \emph{Nature} \textbf{1996},
  \emph{383}, 313--318\relax
\mciteBstWouldAddEndPuncttrue
\mciteSetBstMidEndSepPunct{\mcitedefaultmidpunct}
{\mcitedefaultendpunct}{\mcitedefaultseppunct}\relax
\EndOfBibitem
\bibitem[Clifford and Williams(2000)Clifford, and Williams]{Clifford2000}
Clifford,~A.~A.; Williams,~J.~R. In \emph{Supercritical Fluid Methods and
  Protocols}; Williams,~J.~R., Clifford,~A.~A., Eds.; Humana Press: New Jersey,
  2000; pp 1--16\relax
\mciteBstWouldAddEndPuncttrue
\mciteSetBstMidEndSepPunct{\mcitedefaultmidpunct}
{\mcitedefaultendpunct}{\mcitedefaultseppunct}\relax
\EndOfBibitem
\bibitem[Xu \latin{et~al.}(2005)Xu, Kumar, Buldyrev, Chen, Poole, Sciortino,
  and Stanley]{Xu2005}
Xu,~L.; Kumar,~P.; Buldyrev,~S.~V.; Chen,~S.-H.; Poole,~P.~H.; Sciortino,~F.;
  Stanley,~H.~E. {Relation between the Widom line and the dynamic crossover in
  systems with a liquid – liquid phase transition}. \emph{Proceedings of the
  National Academy of Sciences} \textbf{2005}, \emph{102}, 16558--16562\relax
\mciteBstWouldAddEndPuncttrue
\mciteSetBstMidEndSepPunct{\mcitedefaultmidpunct}
{\mcitedefaultendpunct}{\mcitedefaultseppunct}\relax
\EndOfBibitem
\bibitem[Schienbein and Marx(2018)Schienbein, and Marx]{Schienbein2018}
Schienbein,~P.; Marx,~D. {Investigation concerning the uniqueness of separatrix
  lines separating liquidlike from gaslike regimes deep in the supercritical
  phase of water with a focus on Widom line concepts}. \emph{Physical Review E}
  \textbf{2018}, \emph{98}, 022104\relax
\mciteBstWouldAddEndPuncttrue
\mciteSetBstMidEndSepPunct{\mcitedefaultmidpunct}
{\mcitedefaultendpunct}{\mcitedefaultseppunct}\relax
\EndOfBibitem
\bibitem[Gallo \latin{et~al.}(2014)Gallo, Corradini, and Rovere]{Gallo2014}
Gallo,~P.; Corradini,~D.; Rovere,~M. {Widom line and dynamical crossovers as
  routes to understand supercritical water}. \emph{Nature Communications}
  \textbf{2014}, \emph{5}, 5806\relax
\mciteBstWouldAddEndPuncttrue
\mciteSetBstMidEndSepPunct{\mcitedefaultmidpunct}
{\mcitedefaultendpunct}{\mcitedefaultseppunct}\relax
\EndOfBibitem
\bibitem[Simeoni \latin{et~al.}(2010)Simeoni, Bryk, Gorelli, Krisch, Ruocco,
  Santoro, and Scopigno]{Simeoni2010}
Simeoni,~G.~G.; Bryk,~T.; Gorelli,~F.~A.; Krisch,~M.; Ruocco,~G.; Santoro,~M.;
  Scopigno,~T. {The Widom line as the crossover between liquid-like and
  gas-like behaviour in supercritical fluids}. \emph{Nature Physics}
  \textbf{2010}, \emph{6}, 503--507\relax
\mciteBstWouldAddEndPuncttrue
\mciteSetBstMidEndSepPunct{\mcitedefaultmidpunct}
{\mcitedefaultendpunct}{\mcitedefaultseppunct}\relax
\EndOfBibitem
\bibitem[Sun \latin{et~al.}(2020)Sun, Hastings, Ishikawa, Baron, and
  Monaco]{Sun2020a}
Sun,~P.; Hastings,~J.~B.; Ishikawa,~D.; Baron,~A.~Q.; Monaco,~G. {Two-Component
  Dynamics and the Liquidlike to Gaslike Crossover in Supercritical Water}.
  \emph{Physical Review Letters} \textbf{2020}, \emph{125}, 256001\relax
\mciteBstWouldAddEndPuncttrue
\mciteSetBstMidEndSepPunct{\mcitedefaultmidpunct}
{\mcitedefaultendpunct}{\mcitedefaultseppunct}\relax
\EndOfBibitem
\bibitem[Bencivenga \latin{et~al.}(2007)Bencivenga, Cunsolo, Krisch, Monaco,
  Ruocco, and Sette]{Bencivenga2007a}
Bencivenga,~F.; Cunsolo,~A.; Krisch,~M.; Monaco,~G.; Ruocco,~G.; Sette,~F.
  {High-frequency dynamics of liquid and supercritical water}. \emph{Physical
  Review E} \textbf{2007}, \emph{75}, 051202\relax
\mciteBstWouldAddEndPuncttrue
\mciteSetBstMidEndSepPunct{\mcitedefaultmidpunct}
{\mcitedefaultendpunct}{\mcitedefaultseppunct}\relax
\EndOfBibitem
\bibitem[Abascal and Vega(2005)Abascal, and Vega]{Abascal2005}
Abascal,~J. L.~F.; Vega,~C. {A general purpose model for the condensed phases
  of water: TIP4P/2005}. \emph{The Journal of Chemical Physics} \textbf{2005},
  \emph{123}, 234505\relax
\mciteBstWouldAddEndPuncttrue
\mciteSetBstMidEndSepPunct{\mcitedefaultmidpunct}
{\mcitedefaultendpunct}{\mcitedefaultseppunct}\relax
\EndOfBibitem
\bibitem[Stillinger and Weber(1985)Stillinger, and Weber]{Stillinger1985}
Stillinger,~F.~H.; Weber,~T.~A. {Computer simulation of local order in
  condensed phases of silicon}. \emph{Physical Review B} \textbf{1985},
  \emph{31}, 5262--5271\relax
\mciteBstWouldAddEndPuncttrue
\mciteSetBstMidEndSepPunct{\mcitedefaultmidpunct}
{\mcitedefaultendpunct}{\mcitedefaultseppunct}\relax
\EndOfBibitem
\bibitem[Ward \latin{et~al.}(2012)Ward, Zhou, Wong, Doty, and
  Zimmerman]{Ward2012}
Ward,~D.~K.; Zhou,~X.~W.; Wong,~B.~M.; Doty,~F.~P.; Zimmerman,~J.~A.
  {Analytical bond-order potential for the cadmium telluride binary system}.
  \emph{Physical Review B} \textbf{2012}, \emph{85}, 115206\relax
\mciteBstWouldAddEndPuncttrue
\mciteSetBstMidEndSepPunct{\mcitedefaultmidpunct}
{\mcitedefaultendpunct}{\mcitedefaultseppunct}\relax
\EndOfBibitem
\bibitem[Plimpton(1995)]{Plimpton1995}
Plimpton,~S. {Fast Parallel Algorithms for Short-Range Molecular Dynamics}.
  \emph{Journal of Computational Physics} \textbf{1995}, \emph{117},
  1--19\relax
\mciteBstWouldAddEndPuncttrue
\mciteSetBstMidEndSepPunct{\mcitedefaultmidpunct}
{\mcitedefaultendpunct}{\mcitedefaultseppunct}\relax
\EndOfBibitem
\bibitem[Alejandre \latin{et~al.}(1995)Alejandre, Tildesley, and
  Chapela]{Alejandre1995}
Alejandre,~J.; Tildesley,~D.~J.; Chapela,~G.~A. {Molecular dynamics simulation
  of the orthobaric densities and surface tension of water}. \emph{The Journal
  of Chemical Physics} \textbf{1995}, \emph{102}, 4574--4583\relax
\mciteBstWouldAddEndPuncttrue
\mciteSetBstMidEndSepPunct{\mcitedefaultmidpunct}
{\mcitedefaultendpunct}{\mcitedefaultseppunct}\relax
\EndOfBibitem
\bibitem[Vega and Abascal(2011)Vega, and Abascal]{Vega2011}
Vega,~C.; Abascal,~J. L.~F. {Simulating water with rigid non-polarizable
  models: a general perspective}. \emph{Physical Chemistry Chemical Physics}
  \textbf{2011}, \emph{13}, 19663--19688\relax
\mciteBstWouldAddEndPuncttrue
\mciteSetBstMidEndSepPunct{\mcitedefaultmidpunct}
{\mcitedefaultendpunct}{\mcitedefaultseppunct}\relax
\EndOfBibitem
\bibitem[Makhov and Lewis(2003)Makhov, and Lewis]{Makhov2003}
Makhov,~D.~V.; Lewis,~L.~J. {Isotherms for the liquid-gas phase transition in
  silicon from NPT Monte Carlo simulations}. \emph{Physical Review B}
  \textbf{2003}, \emph{67}, 153202\relax
\mciteBstWouldAddEndPuncttrue
\mciteSetBstMidEndSepPunct{\mcitedefaultmidpunct}
{\mcitedefaultendpunct}{\mcitedefaultseppunct}\relax
\EndOfBibitem
\bibitem[Errington \latin{et~al.}(2003)Errington, Debenedetti, and
  Torquato]{Errington2003}
Errington,~J.~R.; Debenedetti,~P.~G.; Torquato,~S. {Quantification of order in
  the Lennard-Jones system}. \emph{The Journal of Chemical Physics}
  \textbf{2003}, \emph{118}, 2256--2263\relax
\mciteBstWouldAddEndPuncttrue
\mciteSetBstMidEndSepPunct{\mcitedefaultmidpunct}
{\mcitedefaultendpunct}{\mcitedefaultseppunct}\relax
\EndOfBibitem
\bibitem[Vega \latin{et~al.}(2006)Vega, Abascal, and Nezbeda]{Vega2006}
Vega,~C.; Abascal,~J. L.~F.; Nezbeda,~I. {Vapor-liquid equilibria from the
  triple point up to the critical point for the new generation of TIP4P-like
  models: TIP4P/Ew, TIP4P/2005, and TIP4P/ice}. \emph{The Journal of Chemical
  Physics} \textbf{2006}, \emph{125}, 34503\relax
\mciteBstWouldAddEndPuncttrue
\mciteSetBstMidEndSepPunct{\mcitedefaultmidpunct}
{\mcitedefaultendpunct}{\mcitedefaultseppunct}\relax
\EndOfBibitem
\bibitem[Mazhukin \latin{et~al.}(2014)Mazhukin, Shapranov, Koroleva, and
  Rudenko]{Mazhukin2014}
Mazhukin,~V.~I.; Shapranov,~A.~V.; Koroleva,~O.~N.; Rudenko,~A.~V. {Molecular
  dynamics simulation of critical point parameters for silicon}.
  \emph{Mathematica Montisnigri} \textbf{2014}, \emph{31}, 64--77\relax
\mciteBstWouldAddEndPuncttrue
\mciteSetBstMidEndSepPunct{\mcitedefaultmidpunct}
{\mcitedefaultendpunct}{\mcitedefaultseppunct}\relax
\EndOfBibitem
\bibitem[Boon and Yip(1991)Boon, and Yip]{Boon1991}
Boon,~J.~P.; Yip,~S. \emph{{Molecular Hydrodynamics}}; Dover Publications: New
  York, 1991\relax
\mciteBstWouldAddEndPuncttrue
\mciteSetBstMidEndSepPunct{\mcitedefaultmidpunct}
{\mcitedefaultendpunct}{\mcitedefaultseppunct}\relax
\EndOfBibitem
\bibitem[Baron(2020)]{Baron2020}
Baron,~A. Q.~R. In \emph{Synchrotron Light Sources and Free-Electron Lasers},
  2nd ed.; Jaeschke,~E.~J., Khan,~S., Schneider,~J.~R., Hastings,~J.~B., Eds.;
  Springer International Publishing: Cham, 2020; pp 2213--2250\relax
\mciteBstWouldAddEndPuncttrue
\mciteSetBstMidEndSepPunct{\mcitedefaultmidpunct}
{\mcitedefaultendpunct}{\mcitedefaultseppunct}\relax
\EndOfBibitem
\bibitem[Hoyer(2004)]{Hoyer2004}
Hoyer,~P.~O. {Non-negative matrix factorization with sparseness constraints}.
  \emph{Journal of Machine Learning Research} \textbf{2004}, \emph{5},
  1457--1469\relax
\mciteBstWouldAddEndPuncttrue
\mciteSetBstMidEndSepPunct{\mcitedefaultmidpunct}
{\mcitedefaultendpunct}{\mcitedefaultseppunct}\relax
\EndOfBibitem
\bibitem[Pitzer(1939)]{Pitzer1939}
Pitzer,~K.~S. {Corresponding States for Perfect Liquids}. \emph{The Journal of
  Chemical Physics} \textbf{1939}, \emph{7}, 583--590\relax
\mciteBstWouldAddEndPuncttrue
\mciteSetBstMidEndSepPunct{\mcitedefaultmidpunct}
{\mcitedefaultendpunct}{\mcitedefaultseppunct}\relax
\EndOfBibitem
\bibitem[Luzar and Chandler(1996)Luzar, and Chandler]{Luzar1996a}
Luzar,~A.; Chandler,~D. {Hydrogen-bond kinetics in liquid water}. \emph{Nature}
  \textbf{1996}, \emph{379}, 55--57\relax
\mciteBstWouldAddEndPuncttrue
\mciteSetBstMidEndSepPunct{\mcitedefaultmidpunct}
{\mcitedefaultendpunct}{\mcitedefaultseppunct}\relax
\EndOfBibitem
\bibitem[Luzar and Chandler(1996)Luzar, and Chandler]{Luzar1996}
Luzar,~A.; Chandler,~D. {Effect of Environment on Hydrogen Bond Dynamics in
  Liquid Water}. \emph{Physical Review Letters} \textbf{1996}, \emph{76},
  928--931\relax
\mciteBstWouldAddEndPuncttrue
\mciteSetBstMidEndSepPunct{\mcitedefaultmidpunct}
{\mcitedefaultendpunct}{\mcitedefaultseppunct}\relax
\EndOfBibitem
\bibitem[Slattery and Bird(1958)Slattery, and Bird]{Slattery1958}
Slattery,~J.~C.; Bird,~R.~B. {Calculation of the diffusion coefficient of
  dilute gases and of the self-diffusion coefficient of dense gases}.
  \emph{AIChE Journal} \textbf{1958}, \emph{4}, 137--142\relax
\mciteBstWouldAddEndPuncttrue
\mciteSetBstMidEndSepPunct{\mcitedefaultmidpunct}
{\mcitedefaultendpunct}{\mcitedefaultseppunct}\relax
\EndOfBibitem
\bibitem[Chapman and Cowling(1939)Chapman, and Cowling]{Chapman1939}
Chapman,~S.; Cowling,~T.~G. \emph{{The mathematical theory of non-uniform
  gases}}; The University press: Cambridge [Eng.], 1939; pp xxiii, 404 p.\relax
\mciteBstWouldAddEndPunctfalse
\mciteSetBstMidEndSepPunct{\mcitedefaultmidpunct}
{}{\mcitedefaultseppunct}\relax
\EndOfBibitem
\bibitem[Doolittle(1951)]{Doolittle1951}
Doolittle,~A.~K. {Studies in Newtonian Flow. II. The Dependence of the
  Viscosity of Liquids on Free‐Space}. \emph{Journal of Applied Physics}
  \textbf{1951}, \emph{22}, 1471--1475\relax
\mciteBstWouldAddEndPuncttrue
\mciteSetBstMidEndSepPunct{\mcitedefaultmidpunct}
{\mcitedefaultendpunct}{\mcitedefaultseppunct}\relax
\EndOfBibitem
\bibitem[Cohen and Turnbull(1959)Cohen, and Turnbull]{Cohen1959}
Cohen,~M.~H.; Turnbull,~D. {Molecular Transport in Liquids and Glasses}.
  \emph{The Journal of Chemical Physics} \textbf{1959}, \emph{31},
  1164--1169\relax
\mciteBstWouldAddEndPuncttrue
\mciteSetBstMidEndSepPunct{\mcitedefaultmidpunct}
{\mcitedefaultendpunct}{\mcitedefaultseppunct}\relax
\EndOfBibitem
\bibitem[Ha \latin{et~al.}(2018)Ha, Yoon, Tlusty, Jho, and Lee]{Ha2018}
Ha,~M.~Y.; Yoon,~T.~J.; Tlusty,~T.; Jho,~Y.; Lee,~W.~B. {Widom Delta of
  Supercritical Gas–Liquid Coexistence}. \emph{The Journal of Physical
  Chemistry Letters} \textbf{2018}, \emph{9}, 1734--1738\relax
\mciteBstWouldAddEndPuncttrue
\mciteSetBstMidEndSepPunct{\mcitedefaultmidpunct}
{\mcitedefaultendpunct}{\mcitedefaultseppunct}\relax
\EndOfBibitem
\bibitem[Brazhkin \latin{et~al.}(2012)Brazhkin, Fomin, Lyapin, Ryzhov, and
  Trachenko]{Brazhkin2012}
Brazhkin,~V.~V.; Fomin,~Y.~D.; Lyapin,~A.~G.; Ryzhov,~V.~N.; Trachenko,~K. {Two
  liquid states of matter: A dynamic line on a phase diagram}. \emph{Physical
  Review E} \textbf{2012}, \emph{85}, 031203\relax
\mciteBstWouldAddEndPuncttrue
\mciteSetBstMidEndSepPunct{\mcitedefaultmidpunct}
{\mcitedefaultendpunct}{\mcitedefaultseppunct}\relax
\EndOfBibitem
\bibitem[Bolmatov \latin{et~al.}(2015)Bolmatov, Zhernenkov, Zav’yalov,
  Stoupin, Cai, and Cunsolo]{Bolmatov2015a}
Bolmatov,~D.; Zhernenkov,~M.; Zav’yalov,~D.; Stoupin,~S.; Cai,~Y.~Q.;
  Cunsolo,~A. {Revealing the Mechanism of the Viscous-to-Elastic Crossover in
  Liquids}. \emph{The Journal of Physical Chemistry Letters} \textbf{2015},
  \emph{6}, 3048--3053\relax
\mciteBstWouldAddEndPuncttrue
\mciteSetBstMidEndSepPunct{\mcitedefaultmidpunct}
{\mcitedefaultendpunct}{\mcitedefaultseppunct}\relax
\EndOfBibitem
\bibitem[Yang and Lee(1952)Yang, and Lee]{Yang1952}
Yang,~C.~N.; Lee,~T.~D. {Statistical Theory of Equations of State and Phase
  Transitions. I. Theory of Condensation}. \emph{Physical Review}
  \textbf{1952}, \emph{87}, 404--409\relax
\mciteBstWouldAddEndPuncttrue
\mciteSetBstMidEndSepPunct{\mcitedefaultmidpunct}
{\mcitedefaultendpunct}{\mcitedefaultseppunct}\relax
\EndOfBibitem
\bibitem[Lee and Yang(1952)Lee, and Yang]{Lee1952}
Lee,~T.~D.; Yang,~C.~N. {Statistical Theory of Equations of State and Phase
  Transitions. II. Lattice Gas and Ising Model}. \emph{Physical Review}
  \textbf{1952}, \emph{87}, 410--419\relax
\mciteBstWouldAddEndPuncttrue
\mciteSetBstMidEndSepPunct{\mcitedefaultmidpunct}
{\mcitedefaultendpunct}{\mcitedefaultseppunct}\relax
\EndOfBibitem
\end{mcitethebibliography}


\providecommand{\latin}[1]{#1}
\makeatletter
\providecommand{\doi}
  {\begingroup\let\do\@makeother\dospecials
  \catcode`\{=1 \catcode`\}=2 \doi@aux}
\providecommand{\doi@aux}[1]{\endgroup\texttt{#1}}
\makeatother
\providecommand*\mcitethebibliography{\thebibliography}
\csname @ifundefined\endcsname{endmcitethebibliography}
  {\let\endmcitethebibliography\endthebibliography}{}
\begin{mcitethebibliography}{15}
\providecommand*\natexlab[1]{#1}
\providecommand*\mciteSetBstSublistMode[1]{}
\providecommand*\mciteSetBstMaxWidthForm[2]{}
\providecommand*\mciteBstWouldAddEndPuncttrue
  {\def\EndOfBibitem{\unskip.}}
\providecommand*\mciteBstWouldAddEndPunctfalse
  {\let\EndOfBibitem\relax}
\providecommand*\mciteSetBstMidEndSepPunct[3]{}
\providecommand*\mciteSetBstSublistLabelBeginEnd[3]{}
\providecommand*\EndOfBibitem{}
\mciteSetBstSublistMode{f}
\mciteSetBstMaxWidthForm{subitem}{(\alph{mcitesubitemcount})}
\mciteSetBstSublistLabelBeginEnd
  {\mcitemaxwidthsubitemform\space}
  {\relax}
  {\relax}

\bibitem[Ward \latin{et~al.}(2012)Ward, Zhou, Wong, Doty, and
  Zimmerman]{Ward2012}
Ward,~D.~K.; Zhou,~X.~W.; Wong,~B.~M.; Doty,~F.~P.; Zimmerman,~J.~A.
  {Analytical bond-order potential for the cadmium telluride binary system}.
  \emph{Physical Review B} \textbf{2012}, \emph{85}, 115206\relax
\mciteBstWouldAddEndPuncttrue
\mciteSetBstMidEndSepPunct{\mcitedefaultmidpunct}
{\mcitedefaultendpunct}{\mcitedefaultseppunct}\relax
\EndOfBibitem
\bibitem[Humphrey \latin{et~al.}(1996)Humphrey, Dalke, and Schulten]{HUMP96}
Humphrey,~W.; Dalke,~A.; Schulten,~K. {VMD} -- {V}isual {M}olecular {D}ynamics.
  \emph{Journal of Molecular Graphics} \textbf{1996}, \emph{14}, 33--38\relax
\mciteBstWouldAddEndPuncttrue
\mciteSetBstMidEndSepPunct{\mcitedefaultmidpunct}
{\mcitedefaultendpunct}{\mcitedefaultseppunct}\relax
\EndOfBibitem
\bibitem[Budininkas \latin{et~al.}(1968)Budininkas, Edwards, and
  Wahlbeck]{Budininkas1968}
Budininkas,~P.; Edwards,~R.~K.; Wahlbeck,~P.~G. {Dissociation Energies of Group
  VIa Gaseous Homonuclear Diatomic Molecules. III. Tellurium}. \emph{The
  Journal of Chemical Physics} \textbf{1968}, \emph{48}, 2870--2873\relax
\mciteBstWouldAddEndPuncttrue
\mciteSetBstMidEndSepPunct{\mcitedefaultmidpunct}
{\mcitedefaultendpunct}{\mcitedefaultseppunct}\relax
\EndOfBibitem
\bibitem[Alejandre \latin{et~al.}(1995)Alejandre, Tildesley, and
  Chapela]{Alejandre1995}
Alejandre,~J.; Tildesley,~D.~J.; Chapela,~G.~A. {Molecular dynamics simulation
  of the orthobaric densities and surface tension of water}. \emph{The Journal
  of Chemical Physics} \textbf{1995}, \emph{102}, 4574--4583\relax
\mciteBstWouldAddEndPuncttrue
\mciteSetBstMidEndSepPunct{\mcitedefaultmidpunct}
{\mcitedefaultendpunct}{\mcitedefaultseppunct}\relax
\EndOfBibitem
\bibitem[Pelissetto and Vicari(2002)Pelissetto, and Vicari]{Pelissetto2002}
Pelissetto,~A.; Vicari,~E. {Critical phenomena and renormalization-group
  theory}. \emph{Physics Reports} \textbf{2002}, \emph{368}, 549--727\relax
\mciteBstWouldAddEndPuncttrue
\mciteSetBstMidEndSepPunct{\mcitedefaultmidpunct}
{\mcitedefaultendpunct}{\mcitedefaultseppunct}\relax
\EndOfBibitem
\bibitem[Poling \latin{et~al.}(2001)Poling, Prausnitz, and
  O’Connell]{Poling2001-sp}
Poling,~B.~E.; Prausnitz,~J.~M.; O’Connell,~J.~P. \emph{{Properties of Gases
  and Liquids, Fifth Edition}}; McGraw-Hill Education: New York, 2001\relax
\mciteBstWouldAddEndPuncttrue
\mciteSetBstMidEndSepPunct{\mcitedefaultmidpunct}
{\mcitedefaultendpunct}{\mcitedefaultseppunct}\relax
\EndOfBibitem
\bibitem[Sampoli \latin{et~al.}(1997)Sampoli, Ruocco, and Sette]{Sampoli1997}
Sampoli,~M.; Ruocco,~G.; Sette,~F. {Mixing of Longitudinal and Transverse
  Dynamics in Liquid Water}. \emph{Physical Review Letters} \textbf{1997},
  \emph{79}, 1678--1681\relax
\mciteBstWouldAddEndPuncttrue
\mciteSetBstMidEndSepPunct{\mcitedefaultmidpunct}
{\mcitedefaultendpunct}{\mcitedefaultseppunct}\relax
\EndOfBibitem
\bibitem[Sun \latin{et~al.}(2020)Sun, Hastings, Ishikawa, Baron, and
  Monaco]{Sun2020a}
Sun,~P.; Hastings,~J.~B.; Ishikawa,~D.; Baron,~A.~Q.; Monaco,~G. {Two-Component
  Dynamics and the Liquidlike to Gaslike Crossover in Supercritical Water}.
  \emph{Physical Review Letters} \textbf{2020}, \emph{125}, 256001\relax
\mciteBstWouldAddEndPuncttrue
\mciteSetBstMidEndSepPunct{\mcitedefaultmidpunct}
{\mcitedefaultendpunct}{\mcitedefaultseppunct}\relax
\EndOfBibitem
\bibitem[Giordano and Monaco(2010)Giordano, and Monaco]{Giordano2010}
Giordano,~V.~M.; Monaco,~G. {Fingerprints of order and disorder on the
  high-frequency dynamics of liquids}. \emph{Proceedings of the National
  Academy of Sciences} \textbf{2010}, \emph{107}, 21985\relax
\mciteBstWouldAddEndPuncttrue
\mciteSetBstMidEndSepPunct{\mcitedefaultmidpunct}
{\mcitedefaultendpunct}{\mcitedefaultseppunct}\relax
\EndOfBibitem
\bibitem[Matsumoto(2007)]{Matsumoto2007}
Matsumoto,~M. {Relevance of hydrogen bond definitions in liquid water}.
  \emph{The Journal of Chemical Physics} \textbf{2007}, \emph{126},
  054503\relax
\mciteBstWouldAddEndPuncttrue
\mciteSetBstMidEndSepPunct{\mcitedefaultmidpunct}
{\mcitedefaultendpunct}{\mcitedefaultseppunct}\relax
\EndOfBibitem
\bibitem[Kumar \latin{et~al.}(2007)Kumar, Schmidt, and Skinner]{Kumar2007}
Kumar,~R.; Schmidt,~J.~R.; Skinner,~J.~L. {Hydrogen bonding definitions and
  dynamics in liquid water}. \emph{The Journal of Chemical Physics}
  \textbf{2007}, \emph{126}, 204107\relax
\mciteBstWouldAddEndPuncttrue
\mciteSetBstMidEndSepPunct{\mcitedefaultmidpunct}
{\mcitedefaultendpunct}{\mcitedefaultseppunct}\relax
\EndOfBibitem
\bibitem[Strong \latin{et~al.}(2018)Strong, Shi, and Skinner]{Strong2018}
Strong,~S.~E.; Shi,~L.; Skinner,~J.~L. {Percolation in supercritical water: Do
  the Widom and percolation lines coincide?} \emph{The Journal of Chemical
  Physics} \textbf{2018}, \emph{149}, 084504\relax
\mciteBstWouldAddEndPuncttrue
\mciteSetBstMidEndSepPunct{\mcitedefaultmidpunct}
{\mcitedefaultendpunct}{\mcitedefaultseppunct}\relax
\EndOfBibitem
\bibitem[Luzar and Chandler(1996)Luzar, and Chandler]{Luzar1996a}
Luzar,~A.; Chandler,~D. {Hydrogen-bond kinetics in liquid water}. \emph{Nature}
  \textbf{1996}, \emph{379}, 55--57\relax
\mciteBstWouldAddEndPuncttrue
\mciteSetBstMidEndSepPunct{\mcitedefaultmidpunct}
{\mcitedefaultendpunct}{\mcitedefaultseppunct}\relax
\EndOfBibitem
\bibitem[Luzar and Chandler(1996)Luzar, and Chandler]{Luzar1996}
Luzar,~A.; Chandler,~D. {Effect of Environment on Hydrogen Bond Dynamics in
  Liquid Water}. \emph{Physical Review Letters} \textbf{1996}, \emph{76},
  928--931\relax
\mciteBstWouldAddEndPuncttrue
\mciteSetBstMidEndSepPunct{\mcitedefaultmidpunct}
{\mcitedefaultendpunct}{\mcitedefaultseppunct}\relax
\EndOfBibitem
\end{mcitethebibliography}

\end{document}


\maketitle

\section{Tellurium gas phase}
\begin{figure}
    \centering
    \includegraphics[width=0.7\columnwidth]{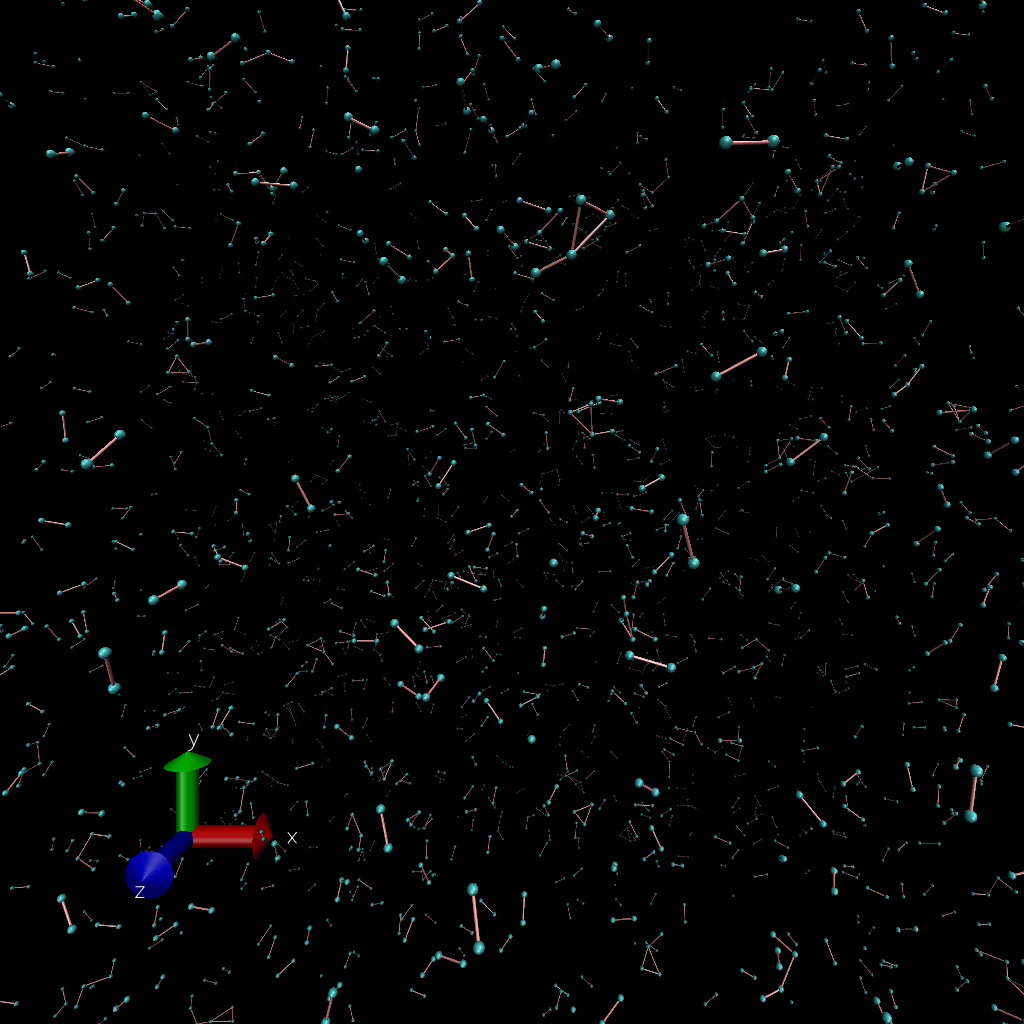}
    \caption{Snapshot of the dilute Te gas state (\SI{100}{bar}, \SI{2760}{K}, \SI{0.109}{g/\cubic\cm}). Atoms are shown as spheres. A bond is shown for each pair of atoms closer than \SI{4.2}{\angstrom}.}
    \label{fig:Te_100bar_2760K_snapshot}
\end{figure}

In the main text we mentioned that the gas phase for the Te bond-order potential (BOP)~\cite{Ward2012} consists of dimer molecules. In Fig.~\ref{fig:Te_100bar_2760K_snapshot} we show a snapshot of a dilute gas state produced using the VMD visualization program~\cite{HUMP96}. A bond is shown for each pair of atoms closer than \SI{4.2}{\angstrom}, same as defined in the main text. We can see that the vast majority of the atoms form dimers; very occasionally, a monomer or a cluster of more than two atoms can be seen. Although it is not the purpose of our work to accurately reproduce all experimental parameters of fluid tellurium, it should be noted that tellurium vapor is known to be diatomic (Te$_2$), with a disassociation energy around \SI{61}{kJ/mol}~\cite{Budininkas1968}.

\begin{figure}
    \centering
    \includegraphics[width=0.7\columnwidth]{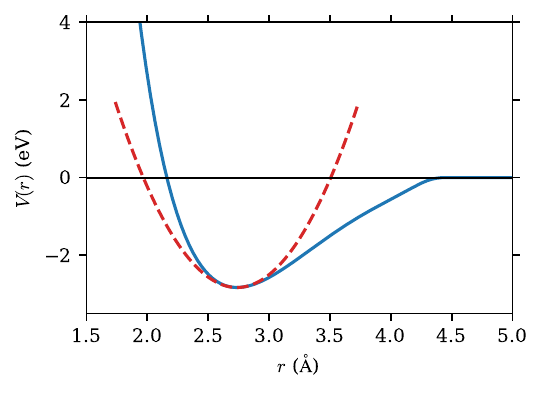}
    \caption{Solid line: potential energy of the Te$_2$ dimer as a function of the distance $r$ between the atoms. Dashed line: harmonic approximation around the potential minimum.}
    \label{fig:Te_pairpotential}
\end{figure}

We can also estimate the dimer oscillation frequency by a harmonic approximation around the minimum of the Te$_2$ pair potential, as shown in Fig.~\ref{fig:Te_pairpotential}. Note that because of the nature of the BOP, this pair potential is only valid when the pair is isolated from other atoms. The potential minimum is at $r=\SI{2.737}{\angstrom}$ with a value of \SI{-2.83}{eV}, or a disassociation energy of \SI{65.3}{kJ/mol}, which is close to the \SI{61}{kJ/mol} value mentioned above. The harmonic approximation is obtained by taking the second derivative of potential at its minimum, giving a value of $k=\SI{9.554}{eV/\square\angstrom}$. Since each Te atom has a mass $m=\SI{127.6}{u}$, the oscillation frequency for the harmonic approximation is $\omega=\sqrt{2k/m}=\SI{38.0}{THz}$, or $\hbar\omega=\SI{25.0}{meV}$, close to the peak position in Fig.~1 in the main text and Fig.~\ref{fig:dispersion} below.

\section{Tellurium critical parameters}
The liquid-gas critical point of the Te BOP has not been reported in the literature. Therefore, we obtain the Te critical parameters using direction MD simulations of the liquid-gas coexistence~\cite{Alejandre1995}. The simulation is done with 1080 atoms in a box with periodic boundary conditions. At each temperature, we first equilibrate using the $NPT$ ensemble at $P=\SI{1000}{bar}$, where the system is in the liquid state. The box size is approximately $\SI{70}{\angstrom} \times \SI{25}{\angstrom} \times \SI{25}{\angstrom}$ after equilibration. Then, the box is expanded in the first dimension by \SI{60}{\angstrom}, and the simulation is run using the $NVT$ ensemble for at least \SI{2}{ns} with a time step of \SI{2.5}{fs}. At each temperature, we fit the density profile along the first dimension with a hyperbolic tangent function for the gas-liquid interface~\cite{Alejandre1995} and obtain the bulk liquid and gas densities. At higher temperatures ($T \geq \SI{1950}{K}$), the boundary is likely to move during the course of the simulation, so we average the density profile every \SI{250}{fs} and fit each average with the hyperbolic tangent method. 

\begin{figure}
    \centering
    \includegraphics[width=0.5\columnwidth]{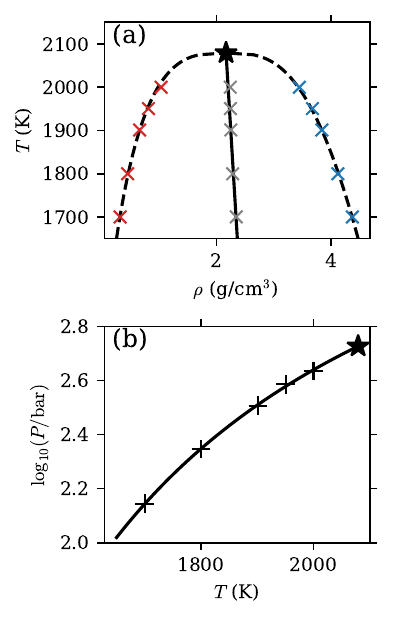}
    \caption{Simulation results for the liquid-gas coexistence curve of Te. (a) Densities of the gas (red crosses) and liquid (blue crosses) phase and their average (grey crosses); also shown are fit results for the law of rectilinear diameters (solid line), scaling of the density difference (dashed line), and critical point (black star). (b) Pressure-temperature data (black ``+'' symbols), best fit using the Antoine equation (solid line), and the critical point (black star).}
    \label{fig:Te_CP}
\end{figure}

The results are shown in Fig.~\ref{fig:Te_CP}a as red and blue crosses representing the densities of the gas phase and liquid phase, respectively. Their average is shown as grey crosses. These data are then fit assuming the law of rectilinear diameters and a scaling exponent for the density difference; i.e., let $\rho_\text{G}(T)$ and $\rho_\text{L}(T)$ represent the densities of the gas and liquid at temperature $T$, then
\begin{eqnarray}
    \frac{1}{2} \left[ \rho_\text{G}(T) + \rho_\text{L}(T) \right] & =& \rho_c + k(T_c-T), \\
    \rho_\text{L}(T) - \rho_\text{G}(T) &=& A \left( \frac{T_c-T}{T_c} \right)^\beta,
\end{eqnarray}
with $\rho_c$, $T_c$, $k$, $A$, and $\beta$ as fit parameters. With fit results are shown in Fig.~\ref{fig:Te_CP}a with the black star indicating the position of the critical point: $\rho_c=\SI[separate-uncertainty = true]{2.17 \pm 0.04}{g/\cubic\cm}$, $T_c=\SI[separate-uncertainty = true]{2080 \pm 40}{K}$. Incidentally, the scaling exponent $\beta = 0.32\pm 0.03$ from the fit is consistent with the 3D Ising exponent of 0.33~\cite{Pelissetto2002}. We then fit the pressure-temperature data shown in Fig.~\ref{fig:Te_CP}b to the Antoine equation~\cite{Poling2001-sp} and get $P_c = \SI[separate-uncertainty = true]{530 \pm 40}{bar}$.

\section{NMF components and dispersion relation}
\begin{figure}
    \centering
    \includegraphics[width=0.7\columnwidth]{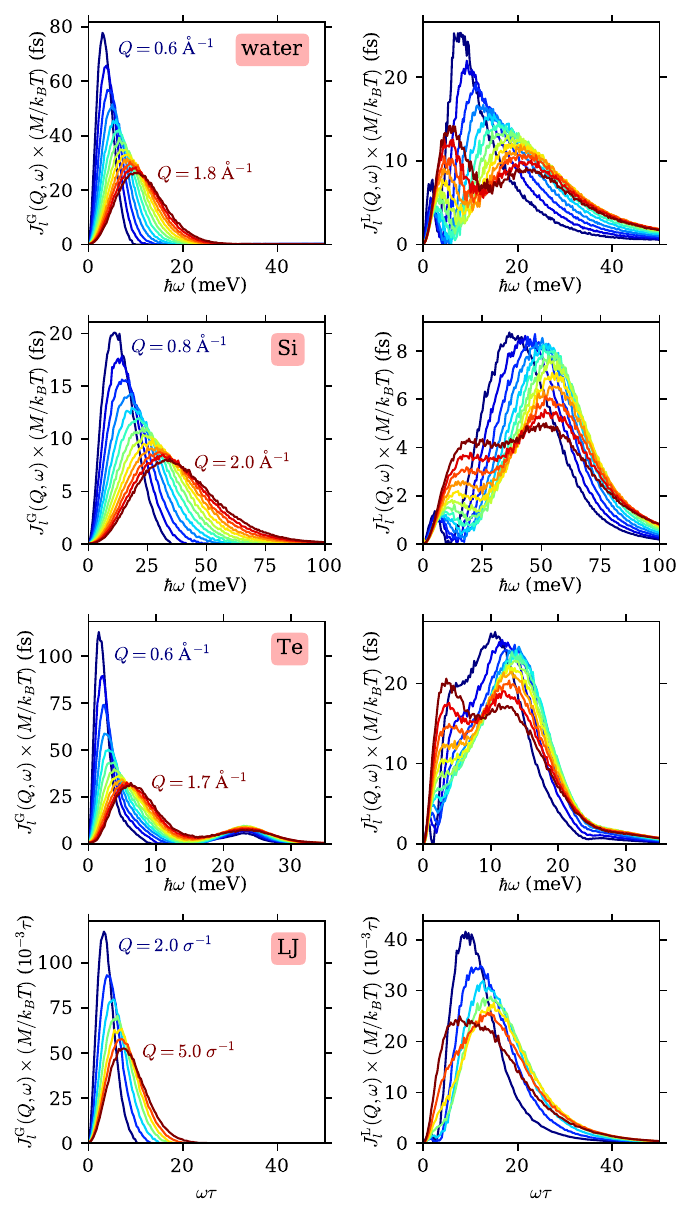}
    \caption{NMF components. Left (right) column shows the G (L) component for each system. From dark blue to dark red is increasing $Q$ with uniform step size; the lowest and highest $Q$ values are annotated in the plot.}
    \label{fig:components}
\end{figure}

\begin{figure}
    \centering
    \includegraphics[width=0.95\columnwidth]{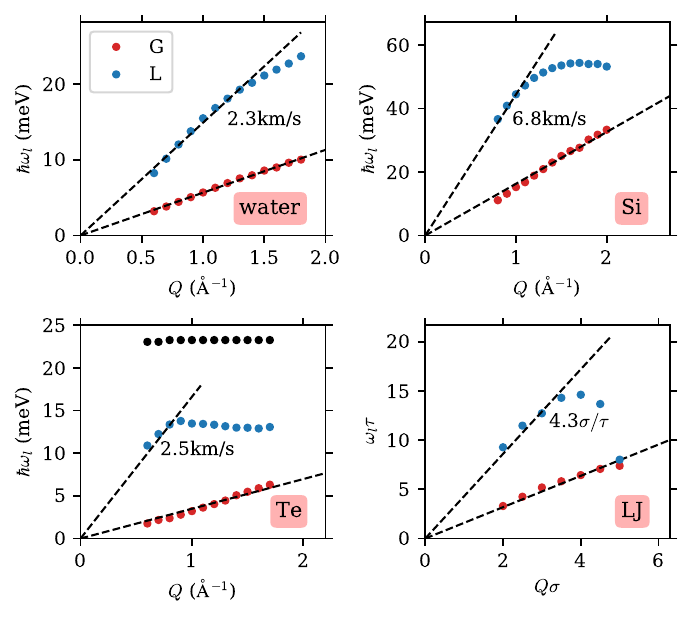}
    \caption{Peak frequency of the G (red) and L (blue) components as function of $Q$. For the L component of water, Si, and Te, only the higher-frequency peak position is plotted (see text). The dashed lines indicate linear relations. The sound speed corresponding to the dispersion relation of the L component is marked on the plot. For Te, the frequency of the dimer oscillation peak in the G component is shown as black dots.}
    \label{fig:dispersion}
\end{figure}

Figure~\ref{fig:components} shows the G and L components obtained from the NMF fit. As expected for spectra in the free-particle limit, the G component shown on the left column has a single Gaussian-like peak which broadens with increasing $Q$ (except for Te with the extra dimer oscillation peak as discussed above). The peak position is expected to be $\sqrt{2}Qv_0 \propto Q$ according to Eq.~(4) in the main text, and this is indeed the case as shown in Fig.~\ref{fig:dispersion}. For Te, we can see in Figs.~\ref{fig:components} and~\ref{fig:dispersion} that the dimer oscillation peak is dispersionless as expected.

The L component exhibits a more interesting behavior as shown on the right column of Fig.~\ref{fig:components}. For water, Si, and Te, the L component appears to contain two peaks in itself, and the one at lower frequency grows with increasing $Q$. This is reminiscent of previous MD simulation results on liquid water~\cite{Sampoli1997} where the two peaks were attributed to transverse and longitudinal modes. Here ``transverse'' and ``longitudinal'' refer to motions with a dominant  transverse and longitudinal polarization, respectively. This is a plausible explanation for the shape of the L component shown here, and we note that the water, Si, and Te potentials all contain terms which give rise to local orientational order. In contrast, the LJ potential is merely pairwise and contains no orientational terms, and its L component does not show two distinct peaks either. However, in all cases, the low-frequency peak in the L component has significant overlap with the G component, making it difficult to distinguish between the two. Thus, the nature of the low-frequency peak cannot easily be determined with the approach used here and may be the subject of future investigations.

The high-frequency peak of the L component, on the other hand, can be directly observed in the $J_l(Q, \omega)$ spectra for all systems, and in the case of water it has been unanimously attributed to the longitudinal acoustic branch~\cite{Sun2020a}. In Fig.~\ref{fig:dispersion} we plot the position of this high-frequency peak as blue dots. It can be seen that the peak frequencies is roughly proportional to $Q$ up to the boundary of the pseudo-Brillouin zone~\cite{Giordano2010,Sun2020a}, which is typical of the behavior of liquids. The sound speeds corresponding to the linear relations are marked on Fig.~\ref{fig:dispersion}. Note that in the case of water, because we have chosen a higher temperature range than in our previous work~\cite{Sun2020a}, the sound speed extracted from the L component is somewhat lower, still fully consistent with the conclusions drawn in this study.

\section{Hydrogen bond definitions}
\begin{figure}
    \centering
    \includegraphics[width=0.95\textwidth]{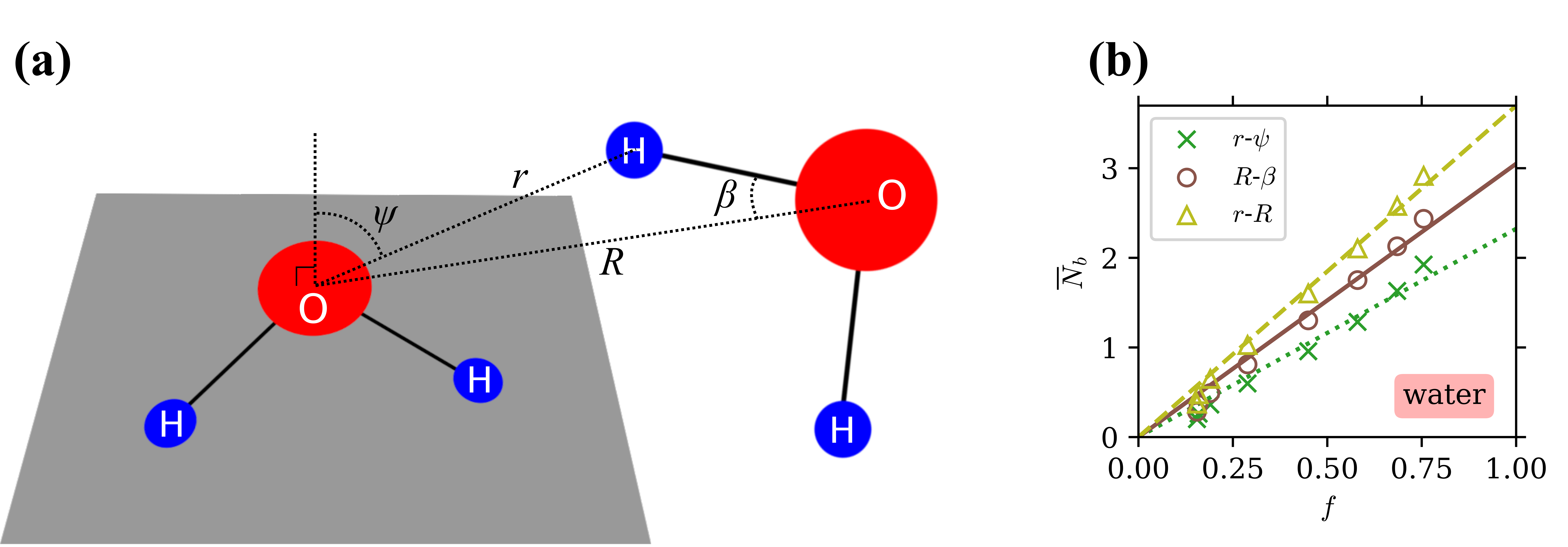}
    \caption{H-bond definitions. Panel (a) shows the variables defined. Panel (b) shows, for each of the three definitions detailed in the text, the average number of hydrogen bonds per molecule, $\overline{N}_b$, plotted again the parameter $f$.}
    \label{fig:water_HBdef}
\end{figure}

A variety of criteria can be found in the literature to define hydrogen-bonding (H-bonding) in water, which result in different H-bond populations~\cite{Matsumoto2007, Kumar2007, Strong2018}. Here, as in our previous work~\cite{Sun2020a}, we adopt the notations in Refs.~\citenum{Kumar2007} and~\citenum{Strong2018}. The variables used are defined in Fig.~\ref{fig:water_HBdef}a. In the main text, we use a definition commonly found in the literature~\cite{Luzar1996a, Luzar1996}: $R < \SI{3.5}{\angstrom}$ and $\beta < \ang{30}$, and we denote it here as the ``$R$-$\beta$'' criterion. Another definition, introduced in Ref.~\citenum{Kumar2007} and denoted here as ``$r$-$\psi$'', requires
\begin{equation}
    (7.1 - 0.050\psi + 0.00021\psi^2) e^{-r/0.343} > 0.0085,
\end{equation}
where $r$ is in units of {\AA}ngstr{\"o}ms and $\psi \in [\ang{0}, \ang{90}]$ in degrees. This turns out to be among the most stringent definitions~\cite{Kumar2007, Strong2018}. A rather relaxed one, denoted here as ``$r$-$R$'', requires $r < \SI{2.5}{\angstrom}$ and $R < \SI{3.5}{\angstrom}$.

In Fig.~\ref{fig:water_HBdef}b we plot the average number of H-bonds per molecule, $\overline{N}_b$, against the parameter $f$ using the definitions mentioned above. In all cases, there is good linearity between $f$ and $\overline{N}_b$, and the data is consistent with an intercept at $\overline{N}_b=0$.

\section{RDF and cutoff for bonding definition}
\begin{figure}
    \centering
    \includegraphics[width=0.95\textwidth]{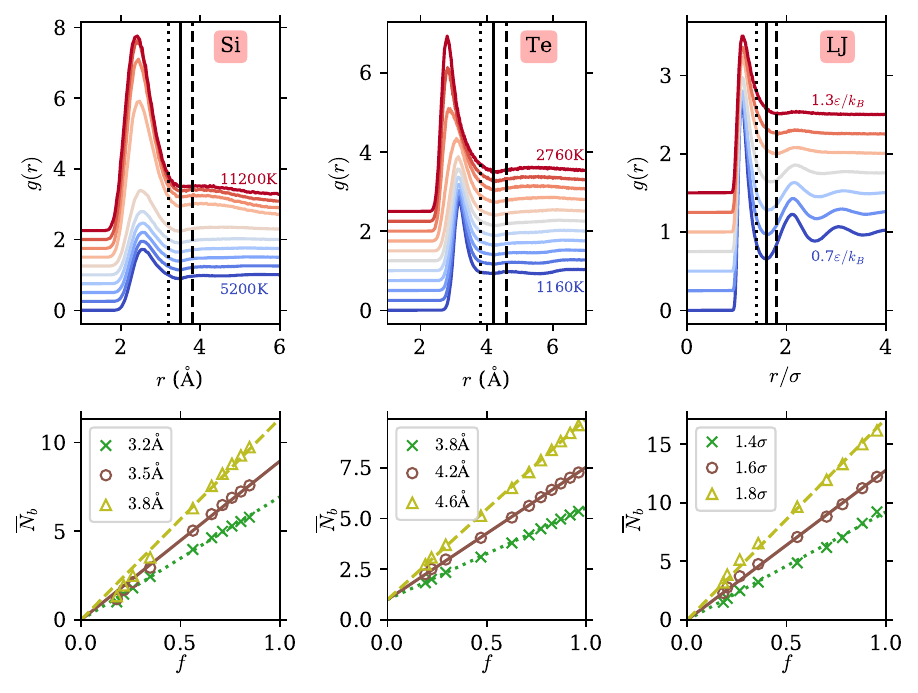}
    \caption{RDF and bonding definitions. Top row: RDF for Si, Te, and LJ fluid along the same isobars ($P=\SI{2850}{bar}$, \SI{870}{bar}, $0.13\varepsilon/\sigma^3$, respectively) and temperature range (annotated on the plot) as in the main text. From dark blue to dark red are increasing temperatures, and an offset of 0.25 is added to $g(r)$ at each temperature step. Vertical lines show the cutoff definitions: solid line is the location of the first minimum in $g(r)$, and the dashed and dotted lines are approximately $\pm 10\%$ away. Bottom row: average number of bonds per atom, $\overline{N}_b$, plotted against the parameter $f$. Symbols represent the data using the cutoff shown in the top row; the exact values are shown in the legend. Lines show linear fits with a fixed intercept ($\overline{N}_b=0$ for Si and LJ, 1 for Te).}
    \label{fig:rcutoff}
\end{figure}

In Fig.~\ref{fig:rcutoff} we show on the top row the radial distribution function (RDF), $g(r)$, for Si, Te, and LJ fluid, on the same isobar and temperature range as in the main text:
\begin{itemize}
    \item Si --- $P=\SI{2850}{bar}$, $T$ from \SI{5200}{K} to \SI{11200}{K} in \SI{600}{K} steps (omitting \SI{10600}{K});
    \item Te --- $P=\SI{870}{bar}$, $T$ from \SI{1160}{K} to \SI{2760}{K} in \SI{160}{K} steps;
    \item LJ --- $P=0.13\varepsilon/\sigma^3$, $T$ from $0.7\varepsilon/k_B$ to $1.3\varepsilon/k_B$ in $0.1\varepsilon/k_B$ steps.
\end{itemize}
The vertical lines show the locations of cutoff distances to define bonding. The solid line corresponds to the definition used in the main text, i.e., the first minimum in $g(r)$ at lower temperatures. In order to show that the conclusions do not depend sensitively on the exact cutoff distance, we also test changing it by $\pm 10\%$, as indicated by the dashed and dotted lines.

On the bottom row, we plot the average number of bonds per atom, $\overline{N}_b$, against the parameter $f$ using the different cutoff definitions. The linearity between $\overline{N}_b$ and $f$ is good for all definitions (perhaps slightly better using the middle one, i.e., at the $g(r)$ minimum). The data are also consistent with an intercept at $\overline{N}_b=0$ for Si and LJ and $\overline{N}_b=1$ for Te, as discussed in the main text.

\bibliography{Supplemental}